\documentclass[12pt]{amsart}

\usepackage{amsfonts,latexsym,amsmath,amssymb, epsfig,array,
delarray}

\title{On an Algebro-Geometric Discretization of KP Hierarchy}
\author{Ali Ulas Ozgur Kisisel}

\newtheorem{Prop}{Proposition}[section]
\newtheorem{Thm}{Theorem}[section]
\newtheorem{Lem}{Lemma}[section]
\newtheorem{Cor}{Corollary}[section]
\newtheorem{Def}{Definition}[section]

\begin{document}

\numberwithin{equation}{section}

\maketitle

\begin{abstract}{This paper studies a certain completely integrable
discretization of the KP hierarchy. This was constructed by
Gieseker in \cite{Gie1}, from certain algebro-geometric data. This
paper has the dual aim of showing that this construction is
generically invertible, and obtaining explicit expressions for the
flow equations. A subsequent article will discuss the Hamiltonian
structure of this system.}
\end{abstract}

\section{The Relation Between the Discretization and Algebro-Geometric Data}
\subsection{Description of the system}
\setcounter{footnote}{0}
\renewcommand{\thefootnote}{}
 \footnote{Acknowledgements: This paper is based on the
author's dissertation at U.C.L.A.. I am indebted to my advisor
David Gieseker for his ideas, guidance and support} We assume that
$N$ and $M$ are positive integers such that $\gcd(N,M)=1$.

Gieseker describes a family of lattice equations parameterized by
$(N,M)$ in \cite{Gie1}. These are two dimensional generalizations
of the periodic Toda lattice equations and reduce to the latter
when $M=1$ . The solutions of the KP equation coming from Riemann
$\Theta$ functions can be approximated by solutions of the lattice
equations, taking $M=N^{2}+1$ and $N\rightarrow \infty$.

Here, we are going to define the system in a roundabout way.
Consider the following problem: We are looking for functions
$\Psi(n,m,t)$ where $(n,m,t) \in \mathbb{Z}/N\mathbb{Z}\times
\mathbb{Z}/M\mathbb{Z}\times \mathbb{C}$ that are almost periodic
in the two space directions of the lattice, i.e.:

\begin{equation}
\begin{split}
\Psi(n+N,m)=\alpha\Psi(n,m) \\ \Psi(n,m+M)=\beta\Psi(n,m)
\end{split}
\end{equation}
where $\alpha, \beta$ are independent of $(n,m)$. Moreover, we
require that $\Psi(n,m+1)$ (suppressing the time variable $t$) is
expressible in terms of three of the $\Psi(k,m)$, more
specifically that $\Psi(n,m+1)$ is of the form:

\begin{equation}
\Psi(n,m+1)=\Psi(n+1,m)-A(n,m)\Psi(n,m)-B(n,m)\Psi(n-1,m)
\end{equation}
where $A(n,m)$ and $B(n,m)$ are periodic in both space entries,
with respective periods $N$ and $M$. Given such a set of $A(n,m),
B(n,m)$ the presence of a nontrivial solution for $\Psi$ forces an
algebraic relation between $\alpha$ and $\beta$. There exists a
matrix $W$ such that the conditions above translate as $\Psi \in
\ker(W)$. To get this $W$, order $\Psi(n,m)$ keeping the second
index more significant than the first (i.e. use the order
$(\Psi(1,1),\Psi(2,1),...$ $\Psi(N,1);\Psi(1,2),...)$). $W$ then
becomes an $NM$ by $NM$ matrix. We present it in block form with
$N$ by $N$ blocks:

\begin{equation} \label{eq:L}
            W = \begin{bmatrix}  -\beta*I_{N}(1) &  0_{N} &   &0_{N}&
X(M)\\
                                     X(1)       & -I_{N}(2) & 0_{N}  &   &
0_{N}\\
                         0_{N}  &    X(2) &   -I_{N}(3)  & 0_{N} &
\\
                            &   &   &   &    \\
                            &...&&...& \\
                            &   &   &   &    \\
                         0_{N}&  &      0_{N}   & X(M-1) & -I_{N}(M)
              \end{bmatrix} \end{equation}
$W$ is in block circulant form. It has two nonzero circulants.
Block $(1,1)$ of $W$ is $-\beta*I_{N}(1)$, and for $i\neq 1$,
block $(i,i)$ is $-I_{N}(i)$. Block $(i+1,i)$ of $W$ is $X(i)$ for
all $i$. (Here, regard $i$ in $\mathbb{Z}/M\mathbb{Z}$.)
$I_{N}(i)$ and $0_{N}$ represent the $N$ by $N$ identity and zero
matrices respectively. The sole purpose of indexing $I_{N}$'s is
making references possible. $X(m)$ is:

\begin{equation}
X(m)=\begin{bmatrix}  -A(1,m)  & 1       &  0  &   & 0 &
-B(1,m)/\alpha
\\
                      -B(2,m)  & -A(2,m) &  1  & 0&    & 0       \\
                         0     & -B(3,m) & -A(3,m)& 1 & 0 &    \\
                               & & & & & \\
                                    &...&   &...&   &...  \\
                               & & & & & \\
                       \alpha  &    0    &     & 0 & -B(N,m) & -A(N,m)
      \end{bmatrix}
\end{equation}
$X(m)$ is a circulant matrix, this time with three nonzero
circulants. The $(i,i)$ entry is $-A(i,m)$, the $(i,i-1)$ entry is
$-B(i,m)$ , and the $(i,i+1)$ entry is $1$ for all $i$ (Here
regard $i$ in $\mathbb{Z}/N\mathbb{Z}$).

We label entries of $W$ with two pairs of numbers. Entry
$((n,m),(k,l))$, where $1\leq n,k \leq N$ and $1\leq m,l \leq M$,
means the entry $(n,k)$ of block $(m,l)$ of $W$. For instance
$W((n,m),(n,m-1))$ is $-A(n,m-1)$, whereas $W((n,m),(n-1,m-1))$ is
$-B(n,m-1)$.

In order for $W\Psi=0$ and $\Psi$ be nontrivial, $\det(W)$ should
be 0. Given a set of $A(n,m), B(n,m)$, this is the defining
equation of a plane algebraic curve in the variables $\alpha$ and
$\beta$. We look at a generic element of this family of curves.
Such a curve has a certain definite behaviour at the loci
$\beta=0, \beta=\infty, \alpha=0$ and $\alpha=\infty$ which we
will see in a moment. Normalize $\Psi$ such that $\Psi(0,0)=1$.
Then, apart from finitely many points of the curve $\det(W)=0$,
$\Psi(n,m)$ is a rational function in $\alpha$ and $\beta$. We
realize each $\Psi(n,m)$ as sections of certain line bundles on
the normalization $\mathcal{X}$ of the curve. Under this
connection, the line bundle corresponding to $\Psi(0,0)$ turns out
to be of degree $g=$genus of the curve. We view this line bundle,
$\mathcal{L}$, as a point of the lifted Jacobian parameterizing
line bundles of degree $g$ on $\mathcal{X}$.

The interesting thing about this correspondence is that it is
possible to move the $A$'s and $B$'s in certain ways keeping the
curve fixed. This can be seen simply by counting dimensions on
either side. On the other hand, under the correspondence, these
degrees of freedom precisely correspond to moving $\mathcal{L}$ in
the Jacobian. From a given curve and a line bundle one can
retrieve $A$'s and $B$'s.

The discrete KP hierarchy is the set of flows corresponding to
moving $\mathcal{L}$ in linear directions on the Jacobian. Since
the curve is fixed for these flows, one can immediately deduce
that there are many conserved quantities: the coefficients of the
curve equation. Since the flows on the Jacobian are linear, the
discrete KP flows have to commute. This is a completely integrable
system. For a discussion of these matters, see \cite{Gie1}.

\subsection{The correspondence}

We now look at the implications of $\det(W)=0$. We are going to
derive most of our results from information about monomials
$f_{i,j}(A,B)\alpha^{i}\beta^{j}$ that appear in $\det(W)$. First
of all, we state the correspondence between the algebro-geometric
data and the discrete KP data. $\mathcal{X}$ denotes the
normalization of the curve $\det{W}=0$.

\begin{Thm}\label{thm:correspondence}
There is a natural correspondence between the following sets of data:

1) A generic smooth curve $\mathcal{X}$ of genus $g$ which
possesses points $P,Q$ such that $N(P-Q)=\mathrm{div}(\alpha)$, an
additional list of points $R_{i},S_{i}$, $i=1,...,M$ so that
$M(P+Q)-\sum(R_{i}+S_{i})=\mathrm{div}(\beta)$, where $\alpha$,
$\beta$ are meromorphic functions on $\mathcal{X}$; and a line
bundle $\mathcal{L}$ of degree $g$ on $\mathcal{X}$ such that

\begin{equation}
H^{0}(X,\mathcal{L}((n+m-1)P+(m-n)Q-\sum_{i=1}^{m}(R_{i}+S_{i}))=0
\end{equation}
for all $(n,m)$.

2) Generic functions $A(n,m),B(n,m)$, periodic in the two space
directions with periods $N$ and $M$.

The relation between $g$, $N$ and $M$ is $g=(N-1)M$. The
$\Psi(n,m)$ obtained from $W\Psi=0$, if properly normalized, are
holomorphic sections of the line bundles $\mathcal{L}((n+m)P
+(m-n)Q-\sum_{i=1}^{m}(R_{i}+S_{i}))$.
\end{Thm}

The construction of (2) from (1), and hence the introduction of
this correspondence is the subject of \cite{Gie1}. It will be
described here later. \cite{Gr} studies the reverse construction
as well, from the point of view of deformation theory. Our
approach is more combinatorial, more in the spirit of \cite{vM-M}.

The equation of the plane curve is $\det(W)=0$. We want to analyze
the behaviour of this curve when $\beta$ or $\alpha$ tend to
$\infty$. We make some linguistic conventions: A ``monomial'' will
mean a full multiplicative expression involving $\alpha$, $\beta$,
$A$, $B$. We distinguish one specific piece of a given monomial:
the ``coefficient'' of the monomial is the piece formed by $A$'s
and $B$'s, in accordance with our treatment of $\alpha$, $\beta$
as variables.

Expand the determinant using all permutations of $n$ letters. A
monomial is said to ``appear'' in the expansion of $\det(W)$ if
there exists a permutation $\pi$ of $NM$ letters so that the
product associated to $\pi$ in the expansion of $\det(W)$ is a
nonzero multiple of this monomial.

\begin{Lem} \label{lem:noncancel}
A monomial appearing in the expansion of $\det(W)$ with a
nonconstant coefficient cannot cancel another monomial with the
same properties in the evaluation of $\det(W)$.
\end{Lem}
\noindent \textbf{Proof:} It will be sufficient to prove that a
monomial with these properties cannot be associated to more than
one permutation. Suppose that $\pi$ is one permutation that such a
monomial $\mathbf{m}$ is associated to. Remove the rows and
columns of $W$ that the $A$'s and $B$'s in $\mathbf{m}$ are on. We
obtain a certain minor $H$ of $W$. Assign all other $A$,$B$'s the
value 0, and call the new matrix $H_{0}$. Clearly, $H_{0}$ has at
most two nonzero elements in each row or column. Suppose that
there is more than one permutation on $H_{0}$ that picks no zero
entries. Then, there exists a row where these two permutations
differ, hence collectively they pick both nonzero entries of that
row. It follows that they also have to pick different entries on
the columns of these entries, and so on. The complete set of
entries picked by one permutation but not the other form at least
one closed loop that can be traversed by changing either the row
or column but not both at one move. But this is impossible.
Indeed, without loss of generality suppose that the diagonal entry
$((b,a),(b,a))$ belongs in such a loop. Then, the only other
nonzero entry of $W$ in this column, $((b-1,a+1),(b,a))$ should
also belong in the loop. Going one step further, the only other
nonzero entry in this new row, $((b-1,a+1),(b-1,a+1))$ should
belong in the loop. Continued, this list contains
$((b-k,a+k),(b-k,a+k))$ for all $k$, which exhausts the diagonal
of $W$ before coming back to the initial point since
$\gcd(N,M)=1$. Therefore $H=W$, which is contradictory to the
assumption that $\pi$ picks at least one $A$ or $B$. $\Box$

The lemma says that the list of $A$,$B$'s in $\mathbf{m}$
determines the associated permutation $\pi$ uniquely, if such a
permutation exists.

\begin{Def}\label{def:degree}
We assign degrees $d$ to multiplicative expressions in
$\alpha,\beta,A,B$ as follows:

\begin{equation}
\begin{split}
 (i)&d(\alpha)=N \\
 (ii)&d(\beta)=M \\
 (iii)&d(A)=1\\
 (iv)&d(B)=2 \\
 (v)&d(c)=0, c\in \mathbb{C}\\
\end{split}
\end{equation}
and the degree of a product is the sum of the degrees.
\end{Def}

The following lemma suggests that this degree assignment is
natural:

\begin{Lem}\label{lem:degree}
If $\mathbf{m}$ is a nonzero monomial appearing in $\det(W)$, then
$d(\mathbf{m})=NM$.
\end{Lem}
\noindent \textbf{Proof:}  According to the definition,
$d((n,1),(n,1))=M$, and $d((n,m),(n,m))=0$ for $m\neq 1$. Observe
that, for nonzero entries of an $X(i)$:

\begin{equation}
  d((n_{1},m),(n_{2},m-1))=n_{1}-n_{2}+1
\end{equation}
Suppose that $\mathbf{m}$ is associated to the permutation $\pi$.
Also, suppose that the number of $\beta$'s that $\pi$ picks is
$k$. Let $S_{1}$ be the multiset (i.e., a collection of elements,
where elements may be listed more than once) of $n_{1}$ such that
$\pi((n_{1},m))=(n_{2},m-1)$ for some value  of $m$ and for some
$n_{2}$. Let $S_{2}$ be the multiset of $n_{2}$'s that appear on
the right hand side of such an equation for some $n_{1}$. If there
are $t$ copies of $n_{1}$ in $S_{1}$, then
$\pi(n_{1},m)=(n_{2},m-1)$ for $t$ values of $m$. But for the
remaining $N-t$ values of $m$, the only possibility that remains
is $\pi(n_{1},m)=(n_{1},m)$. Now, this implies that $n_{1}$
appears on the right hand sides of $N-t$ equations of the latter
form, but then $n_{1}$  satisfies an equation of the form
$\pi((k,m+1))=(n_{1},m)$ for precisely the remaining $t$ values of
$m$. This implies:

\begin{equation}
 S_{1}=S_{2}
\end{equation}
Using this and the previous degree calculations:

\begin{equation}
\begin{split}
 d(\mathbf{m})&=\sum_{n_{1}\in S_{1}}n_{1}-\sum_{n_{2}\in
 S_{2}}n_{2}+M(N-k)+Mk \\
 &=NM
\end{split}
\end{equation}
$\Box$

\begin{Lem}\label{lem:sym}
A monomial of the form  $f(A,B)\alpha^{k}\beta^{j}$ appears in
$\det(W)$ if and only if a monomial of the form
$g(A,B)\alpha^{-k}\beta^{j}$ also does.
\end{Lem}
\noindent \textbf{Proof:} Suppose that $f(A,B)\alpha^{k}\beta^{j}$
is associated to the permutation $\pi$. By Lemma
\ref{lem:noncancel} we know that no combination of monomials in
the expansion cancel. Thus it suffices to display a permutation
that produces a monomial of the form $g(A,B)\alpha^{-k}\beta^{j}$.
We claim that there exists a unique permutation $\pi^{'}$ subject
to the following two conditions: $\pi^{'}$ picks the entry
$((n,m),(n,m))$ of $W$ iff $\pi$ picks the entry
$((n,M-m+1),(n,M-m+1))$, and $\pi^{'}$ picks $((k,m+1),(l,m))$ iff
$\pi^{'}$ picks $((l,M-m+1),(k,M-m))$. It is elementary to verify
that $\pi^{'}$ is a permutation that doesn't pick any zeroes.
Furthermore, $\pi$ and $\pi^{'}$ pick an equal number of elements
from each diagonal block, in particular an equal number of
$\beta$'s. $\pi^{'}$ picks $\alpha^{j}$ from block $(m+1,m)$ iff
$\pi$ picks $\alpha^{-j}$ from block $(M-m+1,M-m)$. Thus $\pi^{'}$
produces a monomial of the form $g(A,B)\alpha^{-k}\beta^{j}$
$\Box$

Notice that, by Lemma \ref{lem:degree}, if a monomial is of the
form $f(A,B)\alpha^{k}\beta^{l}$, then $f(A,B)$ is of degree
$NM-kN-lM$. For another pair of exponents $(k^{'},l^{'})$, suppose
$NM-kN-lM=NM-k^{'}N-l^{'}M$. Then $(k-k^{'})N=(l^{'}-l)M$. Since
$\gcd(N,M)=1$, this implies $N|(l^{'}-l)$. But $0\leq l,l^{'} \leq
N$. Thus either $l=l^{'}$, in which case $k=k^{'}$, or $l^{'}=N$
and $l=0$. But if a permutation picks $N$ $\beta$'s, it has to be
the identity permutation. Thus $k^{'}=0$ and so $k=M$. We conclude
that except for the terms $\alpha^{0}\beta^{N}$ and
$\alpha^{M}\beta^{0}$ which have constant coefficients, the degree
of $f(A,B)$ determines $(k,l)$.

The following Corollary follows from Lemma's \ref{lem:degree} and
\ref{lem:sym}.

\begin{Cor}
A monomial with a coefficient of degree $d$ cannot appear in
$\det(W)$ unless $d$ is among the following list of numbers:

\begin{equation}\label{eq:deg}
\begin{split}
&0\\
&N,N-M,...\\
&2N,2N-M,2N-2M,...\\
&...              \\
&NM,NM-M,NM-2M,NM-3M,...,M,0\\
&...\\
&2NM-2N,2NM-2N-M,2NM-2N-2M,...\\
&2NM-N,2NM-N-M,...\\
&2NM \\
\end{split}
\end{equation}
For $1\leq k\leq M+1$, row $k$ of this list contains the numbers
$(k-1)N-iM$ for $0\leq i \leq \lfloor{\frac{(k-1)N}{M}}\rfloor$
which are all nonnegative. Row $M+1+k$ contains $(M+k)N-iM$ for
$0\leq i \leq \lfloor{\frac{(k-1)N}{M}}\rfloor$, i.e. it has the
same number of entries as row $M+1-k$.
\end{Cor}

Notice that the numbers in $\eqref{eq:deg}$ are situated
symmetrically across the middle row. We make a definition:

\begin{Def} \label{def:i}
Define $i(k)$ to be the number symmetric to $k$ across the middle
row in the list $\eqref{eq:deg}$.
\end{Def}

Notice that $i(k)\equiv k$ mod $2$, since by definition $i(k)-k$ is
$2lN$ where $l$ is the distance between $k$ and the middle row.

$\eqref{eq:deg}$ shows the degrees of coefficients of the
monomials. The corresponding $\alpha^{i}\beta^{j}$ are:

\begin{equation} \label{eq:term}
\begin{split}
&\alpha^{M}\\
&\alpha^{M-1},\alpha^{M-1}\beta^{1},...\\
&\alpha^{M-2},\alpha^{M-2}\beta^{1},\alpha^{M-2}\beta^{2},...\\
&...              \\
&\alpha^{0},\alpha^{0}\beta^{1},\alpha^{0}\beta^{2},\alpha^{0}\beta^{3},...,\alpha^{0}\beta^{N-1},\alpha^{0}\beta^{N}\\
&...\\
&\alpha^{-M+2},\alpha^{-M+2}\beta^{1},\alpha^{-M+2}\beta^{2},...\\
&\alpha^{-M+1},\alpha^{-M+1}\beta^{1},...\\
&\alpha^{-M} \\
\end{split}
\end{equation}
It turns out that each of these terms appear in $\det(W)$ for
generic $A,B$.

We inspect the curve $\det(W)=0$ when $\alpha$ or $\beta$ tend to
$\infty$. $\alpha^{M}$ and $\beta^{N}$ appear in $\det(W)$ with
nonzero constant coefficients. $\alpha^{-M}$ has coefficient
$\Pi_{n,m}B(n,m)$; we assume that $B(n,m)\neq 0$ for any $(n,m)$
so that this coefficient is not zero. If $\alpha$ is finite and
nonzero, it is impossible for $\beta\rightarrow\infty$ on the
curve, since the $\beta^{N}$ term dominates the others.
$\beta=\infty$ implies $\alpha=\infty$ or $\alpha=0$, so there are
two points of the curve at $\beta=\infty$ . Call these points
$\bar{P}$ and $\bar{Q}$ respectively. Similarly, if $\beta$ is
finite it is impossible to have $\alpha=0$ or $\alpha=\infty$.

The next thing we wish to show is that the behaviour of
$\det(W)=0$ at $\bar{P}$ and $\bar{Q}$ is locally identical to the
behaviour of $\alpha^{M}+\beta^{N}=0$ and
$\alpha^{-M}+\beta^{N}=0$ at these points respectively. More
precisely, in the normalization $\mathcal{X}$ of $\det(W)=0$,
there is only one point above either of $\bar{P}$ and $\bar{Q}$
(these will be denoted by $P$ and $Q$). This will imply
$\mathrm{div}(\alpha)=N(P-Q)$, and
$\mathrm{div}_{\infty}(\beta)=M(P+Q)$ where $\alpha, \beta$ on
$\mathcal{X}$ mean composition of $\alpha, \beta$ on $\det(W)=0$
and projection from $\mathcal{X}$ to $\det(W)=0$.

To prove this claim, we present algorithms to blow the curve up at
the points $\bar{P}$ and $\bar{Q}$. These two algorithms are
almost identical, so we explain the procedure for $\bar{Q}$ only.
Set $\hat{\beta}=\frac{1}{\beta}$. So $\bar{Q}$ is the point
$\alpha=\hat{\beta}=0$. To make all exponents nonnegative,
multiply the curve equation by $\alpha^{M}\hat{\beta}^{N}$. The
terms in the curve equation are:

\begin{equation} \label{eq:saldiray1}
\begin{split}
&\alpha^{2M}\hat{\beta}^{N}\\
&\alpha^{2M-1}\hat{\beta}^{N},\alpha^{2M-1}\hat{\beta}^{N-1},...\\
&\alpha^{2M-2}\hat{\beta}^{N},\alpha^{2M-2}\hat{\beta}^{N-1},\alpha^{2M-2}\hat{\beta}^{N-2},...\\
&...              \\
&\alpha^{M}\hat{\beta}^{N},\alpha^{M}\hat{\beta}^{N-1},\alpha^{M}\hat{\beta}^{N-2},\alpha^{M}\hat{\beta}^{N-3},...,\alpha^{M}\hat{\beta}^{1},\boxed{\alpha^{M}}\\
&...\\
&\alpha^{2}\hat{\beta}^{N},\alpha^{2}\hat{\beta}^{N-1},\alpha^{2}\hat{\beta}^{N-2},...\\
&\alpha^{1}\hat{\beta}^{N},\alpha^{1}\hat{\beta}^{N-1},...\\
&\boxed{\hat{\beta}^{N}} \\
\end{split}
\end{equation}
This curve is singular at $\bar{Q}$ iff both $M$ and $N$ are greater than
1. If there is no singularity, there is nothing to prove. So suppose
that the curve is singular at $\bar{Q}$. We blow the curve up at
$\bar{Q}$, set $v=\frac{\alpha}{\hat{\beta}}$ or
$v=\frac{\hat{\beta}}{\alpha}$ depending on whether $N>M$ or $M>N$ in the
respective order. Without loss of generality, we assume that $N>M$.
So we eliminate $\alpha$ from the equation. In the proper transform of
the curve, the terms are:

\begin{equation} \label{eq:saldiray2}
\begin{split}
&v^{2M}\hat{\beta}^{N+M}\\
&v^{2M-1}\hat{\beta}^{N+M-1},v^{2M-1}\hat{\beta}^{N+M-2},...\\
&v^{2M-2}\hat{\beta}^{N+M-2},v^{2M-2}\hat{\beta}^{N+M-3},v^{2M-2}\hat{\beta}^{N+M-4},...\\
&...              \\
&v^{M}\hat{\beta}^{N},v^{M}\hat{\beta}^{N-1},v^{M}\hat{\beta}^{N-2},v^{M}\hat{\beta}^{N-3},...,v^{M}\hat{\beta}^{1},\boxed{v^{M}}\\
&...\\
&v^{2}\hat{\beta}^{N-M+2},v^{2}\hat{\beta}^{N-M+1},v^{2}\hat{\beta}^{N-M},...\\
&v^{1}\hat{\beta}^{N-M+1},v^{1}\hat{\beta}^{N-M},...\\
&\boxed{\hat{\beta}^{N-M}} \\
\end{split}
\end{equation}
Every term except the two boxed terms contain both $v$ and
$\hat{\beta}$ with positive exponents (this will be proven below).
Hence whether or not this blow-up is singular is completely
determined by whether or not $N-M$ or $M$ is 1. Since we assumed
that $M>1$, it is determined by $N-M$ only. If $N-M$ is not $1$,
we blow-up again by setting $w=\frac{v}{\hat{\beta}}$ or
$w=\frac{\hat{\beta}}{v}$ depending on $N-M>M$ or $N-M<M$, and
proceed like this until one of the two exponents in the boxed
terms is one.

The following Lemma explains why we can go on, by establishing
that the two boxes are the deciding terms in every step of the
blowing up.

\begin{Lem}
At any stage of the blow-up algorithm, all of the terms except the
boxed terms (the term on the last row and the term on the
rightmost column) contain both of the variables of that stage with
positive exponents. Therefore whether or not there is a
singularity is completely determined by the exponents of the boxed
terms.
\end{Lem}
\noindent \textbf{Proof:} The proof uses a degree argument. Since
we are using the local coordinate $\hat{\beta}$ rather than
$\beta$, we use a degree definition slightly different from $d$.
Notice, by table $\eqref{eq:deg}$ and $\eqref{eq:saldiray1}$, if
we set $\hat{d}(\alpha)=N$ and $\hat{d}(\hat{\beta})=M$, then the
boxed terms have degree $NM$, and all the other terms have higher
degrees. Whenever a new variable is introduced at some step of the
algorithm, define its degree naturally as the difference of the
degrees of the variables that it is a quotient of. Notice that all
of these degrees are positive because the boxed terms have equal
total degree, and the comparison of exponents ensures that the new
variable is the quotient of the higher degree variable by the
lower degree variable. Furthermore, notice that, with each proper
transform, we are decreasing the degrees of all terms in the list
by a fixed number. So the boxed terms always preserve their
significance of being the only terms of lowest degree. Now suppose
that the variables at some step are $v$ and $w$, boxed terms are
$v^{k}$ and $w^{l}$ (so we should have $k\hat{d}(v)=l\hat{d}(w)$),
and $k>l$. The new variable of the subsequent step is
$y=\frac{w}{v}$ and we eliminate $w$ from the equation. Notice
that $w$'s are replaced by $y$'s with the same exponent, so by
induction they survive with a positive exponent away from the
boxed terms. Contrary to the claim, suppose $v^{i}w^{j}$ is
replaced by $v^{s}y^{j}$ where $s\leq 0$. But then $i+j-l=s\leq
0$. Thus $i+j\leq l$. Since $\hat{d}(v)<\hat{d}(w)$ this implies
that $\hat{d}(v^{i}w^{j})<\hat{d}(w^{l})$. But this is a
contradiction since $w^{l}$ has minimal degree among the terms.
$\Box$

This lemma shows that we may keep blowing up until one of the
exponents of the boxed terms becomes $1$. But then, suppose the
sequence of variables gotten in the blowing ups is
$\alpha,\hat{\beta},v_{1},v_{2},...,v_{k}$. $\hat{\beta}=0$
implies $v_{1}=0$, since in the curve equation at the relevant
stage, all but one term contains $\hat{\beta}$. Repeating this, we
have $v_{2}=0$, $v_{3}=0$, ..., $v_{k}=0$. But the last one is a
nonsingular point, and $v_{k}$ has a simple zero there. Thus there
is one point $Q$ over $\bar{Q}$ in the normalization. Tracing back
from the last step, from the defining relations of each new
variable, it is elementary to see that $\alpha$ has a zero of
order $N$ and $\hat{\beta}$ has a zero of order $M$ at $Q$. As we
remarked, by a similar analysis at $\bar{P}$, there is one point
$P$ above $\bar{P}$ in the normalization of the curve, $\alpha$
has a pole of order $N$, and $\beta$ has a pole of order $M$ at
$P$.

Now, choose a local parameter $z$ at $P$. By the discussion above
$(\alpha,\beta)=(z^{-N}+...,z^{-M}+...)$, where the Laurent expansions are
written starting from the lowest degree terms in $z$ . Similarly, at $Q$,
again in a local parameter $z$, $(\alpha,\beta)=(z^{N}+...,z^{-M}+...)$.
Here, again, the lowest degree terms in the expansion are shown.

We now know the behaviour of the curve at $\beta=\infty$. Notice
that, when $\beta=0$, the curve is defined by the vanishing of a
polynomial of degree $2M$ in $\alpha$. Thus there are $2M$ points
(counted with multiplicity) at $\beta=0$. Call these points
$R_{i},S_{i}$, $i=1,...,M$. We will describe why they are indexed
like this later. It turns out that the $2M$ coefficients of the
equation are algebraically independent, hence almost any set of
$2M$ points can be gotten by an appropriate choice of $A$,$B$.

In order to calculate the genus, we project $\mathcal{X}$ to the
$\beta$ axis, and use the Riemann-Hurwitz formula. The
ramification index at $\beta=\infty$ is $M-1$ at each of the two
points, so the total is $2(M-1)$. Let
$f(\alpha,\beta)=\frac{d}{d\alpha}(\alpha^{M}\det(W))$. Then the
ramification points for finite $\beta$ values are the preimages of
points of intersection of $f=0$ with $\det(W)=0$. We need to
calculate the degree of this divisor only. Since $\det(W)=0$ is
projective, this number is certainly equal to the intersection
number of $\det(W)=0$ and $f=\infty$. $f$ can be $\infty$ only if
$\alpha$ or $\beta$ is $\infty$. The only points of $\det(W)=0$ at
this locus are $\bar{P}$ and $\bar{Q}$. Using the given local
parameters, we inspect the behaviour of $f$ at these points. First
of all, the terms that appear in $f$ are the $\alpha$ derivatives
of the product of $\alpha^{M}$ and the terms in figure
$\eqref{eq:term}$, i.e.

\begin{equation}\label{eq:diffterm}
\begin{split}
&2M\alpha^{2M-1}\\
&(2M-1)\alpha^{2M-2},(2M-1)\alpha^{M-2}\beta^{1},...\\
&(2M-2)\alpha^{2M-3},(2M-2)\alpha^{2M-3}\beta^{1},(2M-2)\alpha^{2M-3}\beta^{2},...\\
&...              \\
&M\alpha^{M-1},M\alpha^{M-1}\beta^{1},M\alpha^{M-1}\beta^{2},M\alpha^{M-1}\beta^{3},...,M\alpha^{M-1}\beta^{N-1},M\alpha^{M-1}\beta^{N}\\
&...\\
&2\alpha^{1},2\alpha^{1}\beta^{1},2\alpha^{1}\beta^{2},...\\
&\alpha^{0},\alpha^{0}\beta^{1},...\\
&0 \\
\end{split}
\end{equation}
Therefore at $P$, the only dominating terms are $\alpha^{2M-1}$
and $\beta^{N}\alpha^{M-1}$, and in the local parameter $z$ both
have Laurent expansions starting with $z^{-N(2M-1)}$. These two
poles cannot cancel each other, because the $\alpha^{2M}$ and
$\alpha^{M}\beta^{N}$ terms cancel in $\det(W)$, but different
constants drop in front with $\alpha$ differentiation.

Similarly, at $Q$, the term $\beta^{N}\alpha^{M-1}$ dominates by
itself, using figure $\eqref{eq:diffterm}$ again. Its local
expansion is $z^{-NM+(M-1)N}+...=z^{-N}+...$.

Therefore, the intersection number of $f=\infty$ and $\det{W}=0$,
which is the total ramification index of $\mathcal{X}$ at finite
points, equals $2MN-N+N=2MN$.

Now we want to show that the projection $\mathcal{X}\rightarrow
(\beta$ axis) for a generic choice of $A$,$B$ does not have any
other branch points. It is enough to show that generic
$\mathcal{X}$ is nonsingular at finite points. We will prove that
$\det(W)=0$ is nonsingular at finite points for generic $A,B$. By
upper semicontinuity theorem, it is sufficient to exhibit one such
curve, and with no new singularities at $\infty$. Let $\eta$ be a
primitive $MN$'th root of unity. Since $\gcd(M,N)=1$, $(1,1)$
generates the additive group
$\mathbb{Z}/N\mathbb{Z}\times\mathbb{Z}/M\mathbb{Z}$. Therefore,
each $(a,a)$ corresponds to a unique representative
$\overline{(a,a)}$ on the toroidal grid for $0\leq a <MN$. Let

\begin{equation} \label{eq:bvalue}
\begin{split}
B(\overline{(a,a)})&=\eta^{a} \\
A(n,m)&=0 \\
\end{split}
\end{equation}
Then, first of all $\Pi_{n,m} B(n,m)=\eta^{s}$ for some $s$. Any
coefficient in the curve equation containing an $A$ vanishes.
Furthermore, suppose that the product
$\mathbf{c}_{(i_{1},j_{1}),...,(i_{k},j_{k})}=B(i_{1},j_{1})...B(i_{k},j_{k})$
appears as a coefficient in some monomial. Since $W$ has toroidal
symmetry, $\mathbf{c}_{(i_{1}+1,j_{1}+1),...,(i_{k}+1,j_{k}+1)}$
also has to appear as part of the coefficient associated to the
same $\alpha^{i}\beta^{j}$. Unless
$\mathbf{c}_{(i_{1},j_{1}),...,(i_{k},j_{k})}$ is the product of
all $B$'s, these two coefficients are different. When $B$'s assume
the values in $\eqref{eq:bvalue}$, the ratio of the second
coefficient to the first is $\eta^{k}$. But
$\sum_{n=1}^{NM}\eta^{kn}=0$ for any $k<NM$. Thus the entire sum
vanishes for such $k$. We deduce that the curve equation for this
assignment of values is:

\begin{equation}\label{eq:nonsingular}
\alpha^{2M}\pm\alpha^{M}\beta^{N}\pm\eta^{s}=0
\end{equation}
It easy to see that $\eqref{eq:nonsingular}$ has no singular points other
than $P$ and $Q$ with the multiplicities described, so the claim is
established.

Therefore, for the generic curve, the total branching is the
calculated minimum, $2NM+2M-2$, not more. By Riemann-Hurwitz formula,

\begin{equation}
\begin{split}
g&=\frac{2NM+2M-2}{2}-2M+1 \\
 &=(N-1)M \\
\end{split}
\end{equation}

Next, we look at the solutions $\Psi$ of $W\Psi=0$. By linear algebra,
the components $\Psi(k,l)$ of $\Psi$ should obey

\begin{equation}\label{eq:ratio}
\frac{\Psi(k,l)}{\Psi(a,b)}=\frac{\det(W^{(i,j),(k,l)})}
{\det(W^{(i,j),(a,b)})}
\end{equation}
for any row $(i,j)$ (recall that we use a pair of indices to
indicate a row or column). Here, $W^{(i,j),(k,l)}$ denotes the
minor of $W$ obtained by removing row $(i,j)$ and column $(k,l)$.

Once and for all, we normalize $\Psi$ such that $\Psi(0,0)=1$.
Then $\Psi(k,l)$ is a meromorphic function on
$\mathcal{X}-\{P,Q\}$ by $\eqref{eq:ratio}$. We first want to
compare the behaviour of the $\Psi(k,l)$'s at the $\beta
-$infinite points $P$ and $Q$ of $\mathcal{X}$. We look at
$\frac{\Psi(k+1,l)}{\Psi(k,l)}$, and
$\frac{\Psi(k,l+1)}{\Psi(k,l)}$. We make a careful choice of
$(i,j)$ for these analyses: Use $(i,j)=(k,l+1)$ for both
comparisons. Repeating the formula above, we have:

\begin{equation}
\frac{\Psi(k+1,l)}{\Psi(k,l)}=\frac
{\det(W^{(k,l+1),(k+1,l)})}
{\det(W^{(k,l+1),(k,l)})}
\end{equation}
and

\begin{equation}
\frac{\Psi(k,l+1)}{\Psi(k,l)}=\frac
{\det(W^{(k,l+1),(k,l+1)})}
{\det(W^{(k,l+1),(k,l)})}
\end{equation}
Although it is definitely difficult to find the determinants on the right
hand side explicitly, it is much easier to determine their
behaviour at $P$ and $Q$. Notice that the element $((k,l+1),(k,l))$ of $W$
is $A(k,l)$. This immediately implies that the
expansion of $\det(W^{(k,l+1),(k,l)})$ contains an $\alpha^{a}\beta^{b}$
term iff the total coefficient of $\alpha^{a}\beta^{b}$ in $\det(W)$
contains $A(k,l)$. The coefficients in $\det(W)$ are toroidally
symmetric, therefore we may deduce that this happens iff the coefficient
of $\alpha^{a}\beta^{b}$ contains any $A$. From this
observation, one can easily make a list of the $\alpha^{a}\beta^{b}$ that
appear in $\det(W^{(k,l+1),(k,l)})$, but the following result will be
enough for us:

\begin{Lem}\label{lem:pole1}
In the expansion of $\det(W^{(k,l+1),(k,l)})$, the terms
$\alpha^{M}$, $\alpha^{-M}$ and $\beta^{N}$ are absent, the terms
with coefficients of degree $1$ and degree $i(1)$ (see definition
\ref{def:i}) are present.
\end{Lem}
\noindent \textbf{Proof:} The absence result is clear. The
coefficients of $\alpha^{M},\alpha^{-M}$ or $\beta^{N}$ never
contain an $A$. For the presence result, note that $1$ and $i(1)$
are both odd numbers. Therefore a monomial consisting in only
$B$'s, which evidently is of even degree, cannot give them. $\Box$

We make a similar analysis for $\det(W^{(k,l+1),(k,l+1)})$. The
element $((k,l+1),(k,l+1))$ of $W$ is a diagonal entry. Hence the
terms that appear in this determinant are precisely those whose
coefficients in the expansion of $\det(W)$ contain a diagonal
entry, and consequently contain a $\beta$. Unless the
$((k,l+1),(k,l+1))$ term is itself $\beta$ (i.e. unless $l=0$) the
complete list of terms in the expansion of
$\det(W^{(k,l+1),(k,l+1)})$ is:

\begin{equation} \label{eq:leftchop}
\begin{split}
&\alpha^{M-1}\beta^{1},...\\
&\alpha^{M-2}\beta^{1},\alpha^{M-2}\beta^{2},...\\
&...              \\
&\alpha^{0}\beta^{1},\alpha^{0}\beta^{2},\alpha^{0}\beta^{3},...,\alpha^{0}\beta^{N-1},\alpha^{0}\beta^{N}\\
&...\\
&\alpha^{-M+2}\beta^{1},\alpha^{-M+2}\beta^{2},...\\
&\alpha^{-M+1}\beta^{1},...\\
\end{split}
\end{equation}

The last determinant that we need to look at is
$\det(W^{(k,l+1),(k+1,l)})$. Similarly, look at terms whose
coefficients in $\det(W)$ contain $((k,l+1),(k+1,l))$. This is an
off-diagonal $1$. Again, we are going to state what we need:

\begin{Lem} \label{lem:pole2}
In the expansion of $det(W^{(k,l+1),(k+1,l)})$, the
$\alpha^{-M}$ and degree $i(1)$ terms are absent, the $\alpha^{M}$
and degree $i(2)$ terms are present.
\end{Lem}
\noindent \textbf{Proof:} The absence of the $\alpha^{-M}$ term
and the presence of $\alpha^{M}$ term are clear. By the symmetry
of Lemma \ref{lem:sym} , an off-diagonal $1$ appears in the degree
$i(1)$ coefficient iff a $B$ appears in the degree $1$
coefficient. This is impossible since $d(B)=2$. Similarly, a $1$
appears in the $i(2)$ coefficient iff a $B$ appears in the degree
$2$ coefficient. This always happens: Pick $B(r,s)$, and now set
all other $A,B=0$ in $W$. Now there are two nonzero elements in
every row and column except the ones that $B(r,s)$ belongs to. The
only nonzero element left in column $(r,s+1)$ is
$((r-1,s+2),(r,s+1))$. Pick this. Continuing, pick all
$((r-t,s+t+1),(r-t+1,s+t))$ until this sequence comes to
$((r-2,s+1),(r-1,s))$. This point appears before
$((r,s+1),(r+1,s))$ in the sequence because $\gcd(N,M)=1$. Pick
all the diagonal elements for the remaining rows and columns. This
gives a valid permutation of the type we want. $\Box$

By using Lemmas \ref{lem:pole1}, \ref{lem:pole2} and figure
$\eqref{eq:leftchop}$, we can compare the behaviour of
different $\Psi(k,l)$ at $P$ and $Q$

\begin{Prop}
$\frac{\Psi(k+1,l)}{\Psi(k,l)}$ has one pole at $P$ and one
zero at $Q$.
\end{Prop}
\noindent \textbf{Proof:} At $P$, the dominating term of
$\det(W^{(k,l+1),(k+1,l)})$ is $\alpha^{M}$. Expanded in the local
parameter, this is $z^{-MN}+...$. The dominating term of
$\det(W^{(k,l+1),(k,l)})$ is the degree $1$ term, and the local
series is $z^{-MN+1}+...$. Thus the claim about $P$ is
established. At $Q$, the dominating terms have degree $i(2)$ and
degree $i(1)$ coefficients respectively. Thus one gets
$z^{-MN+2}+...$ and $z^{-MN+1}+...$ in the local parameter
respectively. This finishes the proof of the claim about $Q$.
$\Box$

\begin{Prop}
$\frac{\Psi(k,l+1)}{\Psi(k,l)}$ has one pole at $P$ and one
pole at $Q$.
\end{Prop}
\noindent \textbf{Proof:} At $P$, the dominating terms of
$\det(W^{(k,l+1),(k,l+1)})$ and  $\det(W^{(k,l+1),(k,l)})$ are
$\beta^{N}$ and the term with degree $1$ coefficient respectively.
Local expansions give $z^{-MN}+...$ and $z^{-MN+1}+...$. And at
$Q$, the dominating terms are $\beta^{N}$ and the term with degree
$i(1)$ coefficient respectively. These give $z^{-MN}+...$ and
$z^{-MN+1}+...$ respectively. $\Box$

So we have obtained that ratios of $\Psi$ obey this fixed
structure of poles at $\beta=\infty$. Now we look at $\beta=0$.
Notice that when $\beta=0$, $\det(W)$ splits as
$\det(X(1))...\det(X(M))$. Likewise, when $\beta=0$ the $\det
(W^{(k,l+1),(k,l)})$ and $\det (W^{(k,l+1),(k+1,l)})$ split as a
product of block determinants. Observe that in these block
expansions, all $\det (X(m))$ appear except for $\det (X(l))$. The
remaining $N-1 \times N-1$ determinant has nonvanishing
determinant at points $R_{i}$, $S_{i}$ of the curve for generic
$A$,$B$. Therefore, these two determinants have the same zeros on
the curve at $\beta=0$. This shows that $\Psi(k,l)$ and
$\Psi(k+1,l)$ have the same divisor at $\beta=0$. On the other
hand, we saw that $\det(L^{(k,l+1),(k,l+1)})$ vanishes identically
at $\beta=0$. Therefore $\Psi(k,l+1)$ has two more zeros than
$\Psi(k,l)$ at $\beta=0$, namely the two solutions for $\alpha$ of
$\det(X(l))$. Call these two points $R_{l}, S_{l}$.

The line bundle mentioned in the correspondence comes from a
divisor $\mathcal{D}$ encoding the totality of remaining poles of
$\Psi$'s:

\begin{Def}
Let $\mathcal{D}$ be a minimal effective divisor such that
\[\mathcal{D}+div(\Psi(n,m))+(n+m)P+(m-n)Q-\sum_{i=1}^{m}(R_{i}+S_{i})
\geq 0\]
for all $(n,m)$.
\end{Def}

We remark that such a finite $\mathcal{D}$ exists: Suppose the
condition is satisfied for $0\leq n<N$ and $0\leq m<M$, then it is
automatically satisfied for all positive $(n,m)$. This happens
since $\Psi(n+kN,m+lM)=\alpha^{k}\beta^{l}\Psi(n,m)$. The new
divisors introduced by the $\alpha$'s and $\beta$'s are exactly
taken care of by the extra points in definition of $\mathcal{D}$.
Define the line bundle $\mathcal{L}=\mathcal{O}(\mathcal{D})$.
Furthermore, set

\begin{equation}
\mathcal{L}_{n,m}=\mathcal{L}((m+n)P+(m-n)Q-\sum_{i=1}^{m}(R_{i}+S_{i}))
\end{equation}
The choice of $\mathcal{D}$ says $\Psi(n,m)$ is a section of
$\mathcal{L}_{n,m}$.

Now, fix $m,n$, and select $k$ large enough so that $deg(\mathcal{D})+k
> 2g-2$. For this particular $k$, Riemann-Roch theorem gives

\begin{equation}
h^{0}(\mathcal{L}_{n,m}(kP))=\deg(\mathcal{L}_{n,m})+k-g+1
\end{equation}
because $h^{1}(\mathcal{L}_{n,m}(kP))=0$  by the choice
of $k$.

\begin{Prop}
$\Psi(n+k,m)$ is a section of
$\mathcal{L}_{n,m}(kP)$
but is not a section of  $\mathcal{L}_{n,m}((k-1)P)$
\end{Prop}
\noindent \textbf{Proof:} $\Psi(n+k,m)$ is a section of
$\mathcal{L}_{n,m}(kP)$ by definition. Suppose it is a section of
$\mathcal{L}_{n,m}((k-1)P)$. Then by the pole comparisons at $P$,
we can replace $k$ by any nonnegative number, and the hypothesis
remains true. But then $\Psi(n,m)$ is a section of
$\mathcal{L}_{n,m}(-P)$, and by pole comparisons again this holds
for all $(n,m)$. In particular $\Psi(0,0)=1$ is a section of
$\mathcal{L}(-P)$, therefore $\mathcal{D}-P$ is effective. But
these imply that we can replace $\mathcal{D}$ by $\mathcal{D}-P$,
which contradicts the minimality of $\mathcal{D}$. $\Box$

From the proposition we deduce that

\begin{equation}
h^{0}(\mathcal{L}_{n,m}(kP))=deg(\mathcal{D})+k-g+1
\end{equation}
for all $k$, in particular
$h^{0}(\mathcal{L}_{n,m})=deg(\mathcal{D})-g+1$. Indeed,
increasing $k$ by one always gives a new section, therefore,
$h^{1}$ should be zero at each step, because it is eventually
zero.

The divisor $\mathcal{D}$ is of degree $g$. For a proof of this,
we refer the reader to \cite{Gr}.


This concludes the construction of the algebro-geometric data from
the difference operator. We now summarize the construction of the
KP data from the algebro-geometric data.

Assume $\mathcal{X}$ is a smooth curve of genus $g$, $P$, $Q$,
$R_{i}$, $S_{i}$ points on $\mathcal{X}$ such that
$\mathrm{div}(\alpha)=N(P-Q)$ and
$\mathrm{div}(\beta)=M(P+Q)-\sum_{i=1}^{M}(R_{i}+S_{i})$,
$\mathcal{L}$ a line bundle of degree $g$ on $\mathcal{X}$ such
that

\begin{equation}
h^{0}(\mathcal{X},\mathcal{L}((n+m-1)P+(m-n)Q-\sum_{i=1}^{m}(R_{i}+S_{i})))=0
\end{equation}
for all $m,n$. Let

\begin{equation}
\mathcal{L}_{n,m}=\mathcal{L}((n+m)P+(m-n)Q-\sum_{i=1}^{m}(R_{i}+S_{i}))
\end{equation}

as before. Then,

\begin{Prop}
$h^{0}(\mathcal{X},\mathcal{L}_{n,m})=1$ for each $n,m$.
\end{Prop}
\noindent \textbf{Proof:} By Riemann-Roch theorem,
$h^{0}(\mathcal{X},\mathcal{L}_{n,m})\geq g+1-g=1$. On the other
hand, it is impossible to have strict inequality because of the
regularity assumption. (Removing one $P$ prohibits at most one
section.)  $\Box$

Say $\Psi(n,m)$ is a nonzero section of $\mathcal{L}_{n,m}$.

\begin{Prop}\label{prop:dependence}
The set of sections
$\{\Psi(n+1,m),\Psi(n,m),\Psi(n-1,m),\Psi(n,m+1)\}$ is linearly dependent.
\end{Prop}
\noindent \textbf{Proof:} Notice that all four of them are
sections of $\mathcal{L}_{n,m}(P+Q)$. On the other hand, by
Riemann-Roch theorem, we get $h^{0}(\mathcal{L}_{n,m}(P+Q))\geq
3$, and by the regularity assumption again, equality has to be
realized. $\Box$

The ratio $f_{n,m}=\frac{\Psi(n,m)}{\Psi(0,0)}$ has a pole of
order precisely $n+m$ at $P$. Suppose $z$ is a uniformizing
parameter at $P$. We can normalize $\Psi(n,m)$ so that
$f_{n,m}z^{n+m}=1$ at $P$. We can also normalize the meromorphic
functions $\alpha$, $\beta$ such that $\alpha=z^{-N}+...$,
$\beta=z^{-M}+...$.

After these normalizations it is elementary to show that the
linear dependence between the four sections in Prop
\ref{prop:dependence} has to be of the form

\begin{equation}
\Psi(n+1,m)-\Psi(n,m+1)-A(n,m)\Psi(n,m)-B(n,m)\Psi(n,m)=0
\end{equation}
for some functions $A$, $B$ on $\mathbb{Z}/N\mathbb{Z}\times
\mathbb{Z}/M\mathbb{Z}$.

Thus we get back to $A$,$B$, having started from the
algebro-geometric data. Next, we wish to calculate the flow
equations explicitly by computing the effect of moving
$\mathcal{L}$ in the Jacobian in certain directions.

\section{The flow equations}

\subsection{Some functions on $ \mathbb{Z}/N\mathbb{Z} \times
\mathbb{Z}/M\mathbb{Z}$} \label{sec:kappa}

As before, suppose that $N,M \in \mathbb{Z}$, and $\gcd(N,M)=1$. Let $S$
denote the set of functions
$f:\mathbb{Z}/N\mathbb{Z} \times \mathbb{Z}/M\mathbb{Z}\rightarrow
\{-1,0,1\}$.

\begin{Prop} \label{prop:kappa}
There is a unique function $\kappa$ in $S$
that satisfies the following conditions:

\begin{equation} \label{eq:kappadiff}
\begin{split}
  (i)&\kappa(0,0)-\kappa(1,-1) =-1 \\
  (ii)&\kappa(0,1)-\kappa(1,0)  =1  \\
  (iii)&\kappa(-1,0)-\kappa(0,-1)=1 \\
  (iv)&\kappa(-1,1)-\kappa(0,0) =-1 \\
\end{split}
\end{equation}
and except for these four values of $(i,j)$, $\kappa(i-1,j+1)=\kappa(i,j)$
\end{Prop}
\noindent \textbf{Proof:} Uniqueness is easy to prove, because if
two such functions exist, their  difference has to be a constant.
But by (i) and (iv) the only possibility for $\kappa(0,0)$ is $0$.
Therefore the constant is zero.

To prove existence, we note
that since $\gcd(N,M)=1$,
$(-1,1)$ is a generator. Look at the sequence

\begin{equation}\label{eq:seq}
(-1,1),(-2,2),...,(-a,a),...
\end{equation}
We will distinguish the two cases below:

1) Suppose in the sequence $\eqref{eq:seq}$, $(1,0)$ appears before
$(-1,0)$. Then we declare

\begin{equation}
\begin{split}
\kappa(-1,1)&=\kappa(-2,2)=...=\kappa(1,0)=-1 \\
\kappa(1,-1)&=\kappa(2,-2)=...=\kappa(-1,0)=1 \\
\end{split}
\end{equation}
and $\kappa(a,b)=0$ if $(a,b)$ is not in these lists.

2) Suppose $(-1,0)$ appears before $(1,0)$. We declare

\begin{equation}
\begin{split}
\kappa(-1,1)&=\kappa(-2,2)=...=\kappa(0,-1)=-1 \\
\kappa(1,-1)&=\kappa(2,-2)=...=\kappa(0,1)=1   \\
\end{split}
\end{equation}
and $\kappa(a,b)=0$ if $(a,b)$ is not in these lists.

One can check that $\kappa$ satisfies the conditions that we asked for.
$\Box$

\noindent \textbf{Remark:} One may wonder which case happens when.
It turns out that the deciding quantity is the parity of the
number of steps in the Euclidean algorithm for the ordered pair
$(N,M)$. In particular we have alternate cases for $(N,M)$ and
$(M,N)$.

Note that $\kappa(n,m)=-\kappa(-n,-m)$.

We introduce another function $\rho$ in $S$:

\begin{Def}\label{def:rhophi}
\begin{gather}
\rho(n,m)=\kappa(n+1,m)+\kappa(n,m) +
\delta_{(n,m),(0,0)}-\delta_{(n,m),(-1,0)} \\
\phi(n,m)=-\rho(-n-1,-m)-\rho(-n,-m)
\end {gather}
\end{Def}
($\delta$ is the Kronecker delta function, i.e. $\delta_{X,Y}=1$
if $X=Y$, and $0$ otherwise, etc.)

Then the following holds:

\begin{equation}
\begin{split}
\rho(n-1,m+1)-\rho(n,m)=&(\kappa(n,m+1)-\kappa(n+1,m))\\
&+(\kappa(n-1,m+1)-\kappa(n,m)) \\
&+\delta_{(n-1,m+1),(0,0)}-\delta_{(n-1,m+1),(-1,0)}\\
&-\delta_{(n,m),(0,0)}
+\delta_{(n,m),(-1,0)}\\
\end{split}
\end{equation}
Therefore $\rho$ is the unique function in $S$ satisfying the following
conditions:
\begin{equation}\label{eq:rhodiff}
\begin{split}
  (i)&\rho(-2,0)-\rho(-1,-1) =1 \\
  (ii)&\rho(0,1)-\rho(1,0)  =1  \\
  (iii)&\rho(-1,0)-\rho(0,-1)=-1 \\
  (iv)&\rho(-1,1)-\rho(0,0) =-1  \\
\end{split}
\end{equation}
and $\rho(n-1,m+1)=\rho(n,m)$ for all other $(n,m)$.

\subsection{Derivation of the discrete KP equations}

We would like to compute what happens to $A$ and $B$ when we keep
$\mathcal{X}$ and the points fixed, but deform the line bundle
$\mathcal{L}$ linearly in the Jacobian. By choosing the direction
of deformation carefully, the computation of the first few
evolution equations for $A$ and $B$ becomes feasible.

$\mathcal{X}$ is mapped inside its Jacobian by the Abel-Jacobi map
.We are going to deform the line bundle in the tangent direction
to the curve at $P$. Assume again that $z$ is a local parameter at
$P$. For small $t\in\mathbb{C}$, define

\begin{equation}
\mathcal{L}_{n,m,t}=\mathcal{L}_{n,m}(P-T)
\end{equation}
where $T=z^{-1}(t)$. First, notice that
$h^{0}(\mathcal{L}_{n,m,t}(-P))=0$ by upper semicontinuity for
small $t$, since $h^{0}(\mathcal{L}_{n,m}(-P))=0$. Thus
$h^{0}(\mathcal{L}_{n,m,t})=1$. Suppose $\Psi(n,m,t)$ is a nonzero
section.

\begin{Prop}
The set $\{\Psi(n,m,t),\Psi(n,m),\Psi(n+1,m)\}$ is
linearly dependent.
\end{Prop}
\noindent \textbf{Proof:} This follows from the observation that
all three are sections of $\mathcal{L}_{n,m,t}(T)$ and
$h^{0}(\mathcal{L}_{n,m,t}(T))=2$ by Riemann-Roch theorem. $\Box$

Thus suppose

\begin{equation}\label{eq:aexpand}
\Psi(n,m,t)(z)=a_{0}(n,m,t)\Psi(n,m)(z)+a_{1}(n,m,t)\Psi(n+1,m)(z)
\end{equation}
where $a_{0}$ and $a_{1}$ are holomorphic in $t$. Now,
$\Psi(n,m,t)$ vanishes at $z=t$ by definition. Therefore

\begin{equation}\label{eq:power}
a_{0}(n,m,t)\Psi(n,m)(t)+a_{1}(n,m,t)\Psi(n+1,m)(t)=0
\end{equation}
Now we can expand this expression in powers of $t$ and calculate the first
few terms of the power series expansions of the holomorphic functions
$a_{i}$ in terms of the Laurent coefficients of the function
$\frac{\Psi(n,m)}{\Psi(0,0)}$ whose pole behaviour at $P$ we know. Let's
give a name to these coefficients.

First define
\begin{equation}
f(n,m)=\frac{\Psi(n,m)}{\Psi(0,0)}
\end{equation}
$f(n,m)$ has a pole of order $n+m$ at $P$. Then let $d_{i}(n,m)$ be such
that

\begin{equation}
f(n,m)=z^{-n-m}+d_{1}(n,m)z^{-n-m+1}+d_{2}(n,m)z^{-n-m+2}+...
\end{equation}

Then since
\begin{equation}
\Psi(n+1,m)=\Psi(n,m+1)+A(n,m)\Psi(n,m)+B(n,m)\Psi(n-1,m)
\end{equation}
we have
\begin{equation}\label{eq:da}
d_{1}(n,m)-d_{1}(n,m+1)=A(n,m)
\end{equation}
Now we can expand $\eqref{eq:power}$ in powers of $t$. We define one more
piece of notation for this purpose: say
$a_{i}(n,m,t)=\sum_{j}a_{i,j}(n,m)t^{j}$.

One sees immediately that $a_{1,1}(n,m)=-1$, therefore we can
normalize  $\Psi(n,m,t)$ in a neighborhood of $P$, replacing it by
$\Psi(n,m,t)t/a_{1}(n,m,t)$.

Then if we compute the coefficient of $t^{-n-m+1}$ we get
\begin{equation}
a_{0,1}(n,m) = d_{1}(n+1,m)-d_{1}(n,m)
\end{equation}
On the other hand, one can expand the relation
\begin{equation}
\Psi(n+1,m,t)=\Psi(n,m+1,t)+A(n,m,t)\Psi(n,m,t)+B(n,m,t)\Psi(n-1,m,t)
\end{equation}
using equation $\eqref{eq:aexpand}$. One can further write all $\Psi(k,l)$
in terms of $\Psi(n+i,m)$ by reducing $l$ to $m$ using the difference
relation repeatedly. Finally, one gets a time dependent linear relation
between $\Psi(k,m)$, where $m$ is fixed. But these $\Psi$ all have
different order poles at $P$ as we have seen, so they are linearly
independent. Therefore many time dependent quantities (that one may
calculate from the last equation) ought to vanish. The
coefficients of $t$ in front of $\Psi(n,m)$ and $\Psi(n-1,m)$ give
evolution equations for $A$ and $B$:

\begin{equation}
\begin{split}
\dot{A}(n,m) &=a_{1,1}(n,m+1)B(n+1,m)+(a_{0,1}(n,m+1)-a_{0,1}(n,m))A(n,m)\\
             &-a_{1,1}(n-1,m)B(n,m)\\
\dot{B}(n,m) &=(a_{0,1}(n,m+1)-a_{0,1}(n-1,m))B(n,m)\\
\end{split}
\end{equation}
Writing $a$'s in terms of $d$'s, equations become
\begin{equation}
   \begin{split}
\dot{A}(n,m) &=B(n,m)-B(n+1,m)+(d_{1}(n+1,m+1)-d_{1}(n,m+1) \\
             &-d_{1}(n+1,m)+d_{1}(n,m))A(n,m)              \\
\dot{B}(n,m) &=(d_{1}(n+1,m+1)-d_{1}(n,m+1)-d_{1}(n,m)      \\
             &+d_{1}(n-1,m))B(n,m) \\
\end{split}
\end{equation}
$d_{1}(0,0)=0$ by definition.
Now, in the light of $\eqref{eq:da}$, the terms in both parentheses are
linear polynomials in $A$'s.
\begin{equation}
\begin{split}
\dot{A}(n,m)=&B(n,m)-B(n+1,m)+(\sum g_{(n,m),(k,l)}A(k,l))A(n,m) \\
\dot{B}(n,m)=&(\sum h_{(n,m),(k,l)} A(k,l))B(n,m)  \\
\end{split}
\end{equation}
By toroidal symmetry,
\begin{equation}
g_{(n+a,m+b),(k+a,l+b)}=g_{(n,m),(k,l)}
\end{equation}
for any $(a,b)\in \mathbb{Z}/N\mathbb{Z}\times\mathbb{Z}/M\mathbb{Z}$.
Same result holds for $h$. Thus, with some abuse of notation, we may write
\begin{equation}\label{eq:arbitrary}
\begin{split}
\dot{A}(n,m)=& B(n,m)-B(n+1,m)+(\sum g(k-n,l-m)A(k,l))A(n,m) \\
\dot{B}(n,m)=& (\sum h(k-n,l-m)A(k,l))B(n,m)  \\
\end{split}
\end{equation}
We subtract the terms of the first equation of $\eqref{eq:arbitrary}$ in parentheses for $(n,m)$
from the same terms for $(n+1,m-1)$:
\begin{equation}
\begin{split}
& (\sum g(k-n-1,l-m+1)A(k,l))-(\sum g(k-n,l-m)A(k,l)) =  \\
& (d_{1}(n+2,m)-d_{1}(n+1,m)-d_{1}(n+2,m-1)+d_{1}(n+1,m-1)) \\
& -(d_{1}(n+1,m+1)-d_{1}(n,m+1)-d_{1}(n+1,m)+d_{1}(n,m)) \\
& = A(n+1,m)-A(n,m)-A(n+1,m-1)+A(n,m-1)  \\
\end{split}
\end{equation}
And we do the same for the second equation:
\begin{equation}
\begin{split}
& (\sum h(k-n-1,l-m+1)A(k,l))-(\sum h(k-n,l-m)A(k,l)) = \\
& (d_{1}(n+2,m)-d_{1}(n+1,m)-d_{1}(n+1,m-1)+d_{1}(n,m-1)) \\
& -(d_{1}(n+1,m+1)-d_{1}(n,m+1)-d_{1}(n,m)+d_{1}(n-1,m)) \\
& = A(n+1,m)-A(n,m)-A(n,m-1)+A(n-1,m-1)  \\
\end{split}
\end{equation}
Equating the coefficients for all $(k,l)$ we obtain
\begin{equation}
\begin{split}
(i)&g(0,1)-g(1,0) =1  \\
(ii)&g(-1,1)-g(0,0) =-1 \\
(iii)&g(0,0)-g(1,-1) =-1 \\
(iv)&g(-1,0)-g(0,-1) =1 \\
\end{split}
\end{equation}
and $g(n+1,m-1)=g(n,m)$ for any other $(n,m)$.

and

\begin{equation}
\begin{split}
(i)&h(0,1)-h(1,0)=1 \\
(ii)&h(-1,1)-h(0,0)=-1 \\
(iii)&h(-1,0)-h(0,-1)=-1 \\
(iv)&h(-2,0)-h(-1,-1)=1 \\
\end{split}
\end{equation}
and $h(n+1,m-1)=h(n,m)$ for any other $(n,m)$

These are precisely the relations for $\kappa$ and $\rho$, so we have
proven:

\begin{Prop}
The equations of evolution are
\begin{equation}
\begin{split}\label{eq:evolution}
 \dot{A}(n,m) =& B(n,m)-B(n+1,m)+(\sum\kappa(k-n,l-m)A(k,l))A(n,m) \\
 \dot{B}(n,m) =& (\sum \rho(k-n,l-m)A(k,l))B(n,m) \\
\end{split}
\end{equation}
\end{Prop}

\end{document}